\documentclass[11pt]{article}
\usepackage[preprint]{acl}
\usepackage{times}
\usepackage{listings}
\usepackage[table]{xcolor}
\usepackage{latexsym}
\usepackage[T1]{fontenc}
\usepackage[utf8]{inputenc}
\usepackage{microtype}
\usepackage{inconsolata}
\usepackage{graphicx}
\usepackage{booktabs}
\usepackage{amsmath}
\usepackage{amssymb}
\usepackage{array}
\usepackage{url}
\usepackage{pifont}   
\usepackage{hyperref}   

\usepackage{booktabs}
\usepackage{multirow}

\hypersetup{
    colorlinks=true,
    linkcolor=magenta,
    citecolor=magenta,
    urlcolor=magenta,
}
\usepackage{tikz}
\usetikzlibrary{arrows.meta, positioning, shapes.geometric, calc}
\usepackage{xspace}

\newcommand{\ourmethod}{\textsc{ExAtlas}\xspace}

\newcommand{\para}[1]{\noindent\textbf{#1}}

\AtBeginDocument{
\setlength{\abovedisplayskip}{6pt}
\setlength{\belowdisplayskip}{6pt}
\setlength{\abovedisplayshortskip}{4pt}
\setlength{\belowdisplayshortskip}{4pt}
}

\title{Building an Atlas of Social Experiments to\\Link Studies, Reconcile Conflicts, and Bridge Gaps}

\author{
  Jiawei Zhang$^{1,=}$ \quad
  Honglin Bao$^{1,=}$ \quad
  Pengda Wang$^{2,=}$ \quad
  Alex Yan$^{3,=}$ \quad \\
  \textbf{Xiao Liu}$^{1,}$\textsuperscript{\ding{41}}  \quad
  \textbf{James A.~Evans}$^{1}$\\
  $^{1}$University of Chicago \quad
  $^{2}$Rice University \quad
  $^{3}$Yale University \\
  $^{=}$ Co-first authors: J.Z., H.B., P.W., and A.Y.\\
  \textsuperscript{\ding{41}} Correspondence: \href{mailto:liuxiao@uchicago.edu}{liuxiao@uchicago.edu}
  \vspace{0.5em}
}

\begin{document}
\maketitle

\begin{abstract}
Social and behavioral science runs thousands of experiments each year, yet their findings rarely accumulate into a coherent map of what is known, what conflicts, and what remains missing. 
We introduce \ourmethod, a framework for turning an archive of experiments into an atlas: a structured map in which studies link, conflict, or leave bridgeable gaps.
Given a target study, \ourmethod searches for prior studies that are locally close in treatment and outcome space and asks whether their observed effects can be composed to predict the target effect. 
This yields three cases. If composition succeeds and agrees with the observed result, \ourmethod \textbf{links} the target to consistent prior evidence. 
If composition succeeds but disagrees, \ourmethod \textbf{reconciles} the conflict and proposes candidate moderators or higher-level theories that could explain it. 
If the composition fails, \ourmethod proposes \textbf{bridge} experiments to close the gap. 
We provide an error bound for composition under local smoothness of the treatment-effect surface. 
On held-out targets certified as locally supported, \ourmethod recovers the direction of the reported causal effect in 98\% of cases across two experimental domains, with an average Spearman correlation of \(\rho=0.770\). 
Human evaluations further suggest that its proposed bridge experiments are plausible and exhibit connectedness, and that its conflict explanations are useful for theory generation.
These results suggest that the archive of social experiments contains more latent structure than current practice extracts—and that making this structure explicit can guide both future theory and future experimentation.

\end{abstract}

\section{Introduction}
\begin{figure*}[th]
    \centering
    \includegraphics[width=1\linewidth]{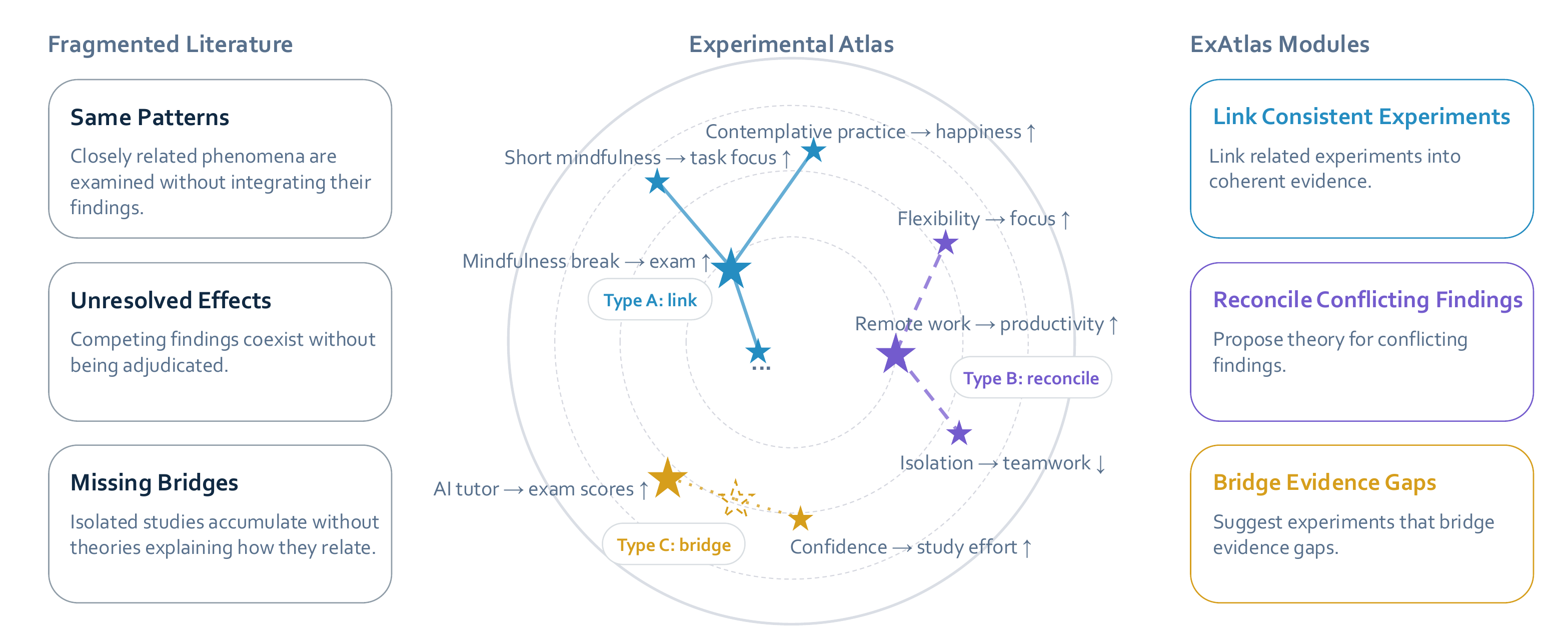}
    \caption{Illustration of how \ourmethod links, reconciles, and bridges experimental evidence.}
    \label{fig-intro}
\end{figure*}

Social and behavioral science has long suffered what Watts called an ``incoherency problem'': scholars examine related phenomena without integrating them, competing findings coexist without being adjudicated, and isolated studies accumulate without theories that explain how they relate \citep{watts2017should}.
Research often proceeds one experiment at a time, leaving the literature fragmented even after decades of productive research~\citep{almaatouq2024beyond}.

Addressing this problem requires turning fragmented experimental evidence into a coherent research landscape. Such a landscape should make three things visible: which experiments speak to the same phenomenon, where they reinforce or contradict each other, and where gaps between them lie. Done well, it serves not as a summary but as an atlas---a map to guide the next theory and the next experiment. Figure~\ref{fig-intro} illustrates this vision: an experimental atlas links consistent experiments, reconciles conflicts, and bridges gaps in fragmented literature. We use \textit{atlas} literally: a structured collection of local maps that together cover a territory, where each map is a region of treatment-outcome space in which experiments compose to predict each other's effects, and where seams between maps are themselves objects of study.

The organization of an experimental landscape is inherently compositional. 
A target finding may bear little surface resemblance to any single prior study, yet be clearly situated within the literature when several studies are interpreted together. 
Consequently, matching experiments by surface similarity alone is insufficient. Directly prompting an LLM does not address the problem either.
What is required is the ability to reason across many studies at once: to detect compositions, conflicts, and gaps, and to determine how the evidential landscape changes when studies are evaluated jointly rather than in isolation. 

We introduce \ourmethod, a framework for constructing such an atlas from joint experimental evidence. 
At the core of \ourmethod{} is a composition step that asks whether a target experiment can be reconstructed from prior observed experiments. Consider a target study asking whether a 15-minute mindfulness break before an exam improves performance for first-generation college students. No single prior experiment may study this exact treatment-outcome pair. But the archive may contain studies of shorter mindfulness interventions before tasks, of study-skills supports targeted at first-generation students, and of brief contemplative practices in academic settings. If the target sits inside the region these studies span, its effect can be predicted by an appropriately weighted average of theirs.


The method is rejectable by design. Theoretically, the composition error decomposes into three terms: extrapolation (how far the target sits from its reconstruction), local model bias (how curved the treatment-effect surface is in the neighborhood), and unobserved residual (variation in factors no embedding captures). When all three are small, the composition is trustworthy. 
Empirically, we evaluate \ourmethod by holding out experiments from an archive and reconstructing their observed treatment effects from the remaining studies. 
On targets that \ourmethod certifies as locally supported, the predicted direction of the causal effect matches the observed direction in 98\% of experiments across management, psychology, and education studies, with an average Spearman correlation of \(\rho=0.770\).

Based on the composition result, \ourmethod identifies conflicts and gaps. When a target is composable from nearby experiments, but the composition disagrees with the observed effect, \ourmethod treats the disagreement as a theoretical opportunity. Prior work and the new result imply different effects despite being close in treatment-outcome space, and an LLM is prompted with both to propose a reconciliation that could explain the divergence. When a target cannot be composed, the gap itself is the finding. \ourmethod iteratively proposes bridge experiments until the target enters the composable region. In this way, the same machinery that predicts known effects also distinguishes two cases: Kuhnian anomalies \citep{Kuhn1962-KUHTSO-3}, which are existing puzzles in the literature where experiments can be composed but yield conflicting predictions, and Lakatosian extensions \citep{lakatos1978methodology}, which mark the field’s next frontiers where existing experiments cannot yet be composed.


Human raters judge \ourmethod's bridge suggestions significantly more plausible and more connected to the source experiments than forced composition of existing experiments. 
This suggests that the bridge module can turn gaps in the archive into actionable experiment designs. 
For composable targets with conflicting evidence, qualitative expert evaluation further shows that \ourmethod can turn disagreements into candidate theoretical explanations. 
For example, \ourmethod surfaces a hidden \(2\times2\) moderator structure in which the effect of candidate age changes direction across cells, and the LLM articulates a reconciliation that experts had not identified but judge plausible and useful for interpreting the conflict.

Our contributions are as follows: 
(1) We introduce \ourmethod, a framework for composing experimental evidence, whose composition procedure turns experimental archives into an atlas of links, conflicts, and gaps. 
(2) We provide a theoretical error bound for composition under local smoothness, and show empirically that when \ourmethod certifies local support, composed effects recover the observed effect direction in 98\% of cases. 
(3) We show that the atlas's two failure modes (i.e., conflicts and gaps) are themselves generative. The first yields theoretical proposals that human raters find substantive, the second yields bridge-experiment proposals that human raters find feasible, plausible, and well connected to the source literature.
\section{Related Work}

\para{LLMs for literature understanding and scientific discovery.}
A growing line of work asks LLMs to generate research ideas and hypotheses from the literature \citep{liu2026heuristicbasedideationfor,yang2024large,baek2025researchagent}, with \citet{si2025can} finding that LLM-generated NLP ideas can match or exceed human experts in novelty. This direction has been extended to experiment design~\citep{li2025agentexpt,guo2026scriptstageautomatingexperimental} and end-to-end research automation~\citep{lu2026towards}.

Most of these methods, however, retrieve literature shallowly, relying on keyword search and feeding titles or abstracts as background, which limits reasoning over relationships among studies. Some work organizes literature more structurally---\citet{baek2025researchagent} draws on academic graphs and entity stores, and \citet{li-etal-2025-chain-ideas} chains ideas across prior work---but the focus is on connecting and extrapolating existing ideas rather than surfacing the conflicts and gaps that often drive scientific progress \citep{Kuhn1962-KUHTSO-3}.
In contrast, \ourmethod constructs an experimental atlas that explicitly models how studies compose, conflict, or leave gaps, and uses these relations to guide theory generation and experiment design.

\para{LLMs as experiment predictors.}
Recent work asks whether LLMs can replicate human experiments by simulating participants in social, economic, psychological, and management studies \citep{aher2023using,horton2023homo,dillion2023can,cui2025large}, with further efforts training models for behavioral simulation \citep{binz2025foundation,cao2025specializing}. A countercurrent questions their reliability for human psychology and policy-relevant inference, particularly in high-stakes domains \citep{schroder2025large,houcan}.

A related line uses LLMs to predict experimental results directly~\citep{hewitt2024predicting,lippert2024can}, treating the model as an outcome predictor that infers participant behavior or treatment effects from an experiment description. These predictions, however, rest on the model's internal knowledge. In contrast, \ourmethod estimates a target effect from existing experiments, and only when those experiments can compose the target, making the evidential basis explicit and the method's applicable scenarios observable.
\section{The \ourmethod Framework}
\label{sec:methods}
\begin{figure*}[t]
\centering
\definecolor{exBlue}{HTML}{1E88C8}
\definecolor{exPurple}{HTML}{6D5BD0}
\definecolor{exYellow}{HTML}{D89B21}
\resizebox{0.92\textwidth}{!}{%
\begin{tikzpicture}[
    node distance=6mm and 9mm,
    >=Latex,
    font=\scriptsize,
    every node/.style={align=center},
    input/.style={
        rectangle,
        rounded corners=2mm,
        draw=black!45,
        line width=0.45pt,
        minimum width=2.55cm,
        minimum height=0.68cm,
        inner sep=3pt,
        fill=black!4
    },
    atlas/.style={
        rectangle,
        rounded corners=2.5mm,
        draw=black!55,
        line width=0.6pt,
        minimum width=2.85cm,
        minimum height=0.84cm,
        inner sep=4pt,
        fill=black!6
    },
    gate/.style={
        diamond,
        aspect=2.45,
        draw=black!50,
        line width=0.5pt,
        minimum width=3.05cm,
        minimum height=0.78cm,
        inner sep=1pt,
        fill=black!5
    },
    outputbase/.style={
        rectangle,
        rounded corners=2mm,
        line width=0.55pt,
        text width=3.6cm,
        minimum height=0.78cm,
        inner sep=3pt
    },
    linkout/.style={outputbase, draw=exBlue, fill=exBlue!9},
    reconcileout/.style={outputbase, draw=exPurple, fill=exPurple!9},
    bridgeout/.style={outputbase, draw=exYellow, fill=exYellow!12},
    line/.style={draw=black!45, line width=0.5pt, -Latex},
    route/.style={draw=black!40, line width=0.45pt, -Latex},
    lab/.style={font=\scriptsize, fill=white, inner sep=1pt, text=black!55}
]

\node[atlas] (atlas) {\ourmethod\\[-1pt]\textit{compose target experiment}};
\node[input, above=of atlas] (archive) {Experimental archive};
\node[input, below=of atlas] (target) {Target experiment};
\node[gate, right=13mm of atlas] (support) {Composable?};

\node[gate, right=8mm of support] (match) {Estimate matches\\observed result?};
\node[bridgeout] (bridge) at (support |- target){Bridge\\[-1pt]\textit{suggest connecting experiments}};
\node[linkout, right=10mm of match, yshift=7mm] (link) {Link\\[-1pt]\textit{connect consistent experiments}};
\node[reconcileout, right=10mm of match, yshift=-7mm] (reconcile) {Reconcile\\[-1pt]\textit{propose competing theory}};

\draw[line] (archive.south) -- (atlas.north);
\draw[line] (target.north) -- (atlas.south);
\draw[line] (atlas) -- (support);
\draw[route] (support) -- node[lab, above] {yes} (match);
\draw[route] (support.south) -- node[lab, right, pos=.45] {no} (bridge.north);
\draw[route] (match.north east) -- node[lab, above, pos=.45] {yes} (link.west);
\draw[route] (match.south east) -- node[lab, below, pos=.45] {no} (reconcile.west);

\end{tikzpicture}%
}
\caption{Overview of the \ourmethod framework. Given a target experiment and an archive of prior experiments, \ourmethod tries to compose the target's treatment and outcome from prior experiments. If the target is composable, \ourmethod estimates its effect from prior observed effects. Consistent estimates link experiments, mismatches motivate theoretical reconciliations, and non-composable targets trigger connecting experiments.}
\label{fig:pipeline}
\end{figure*}
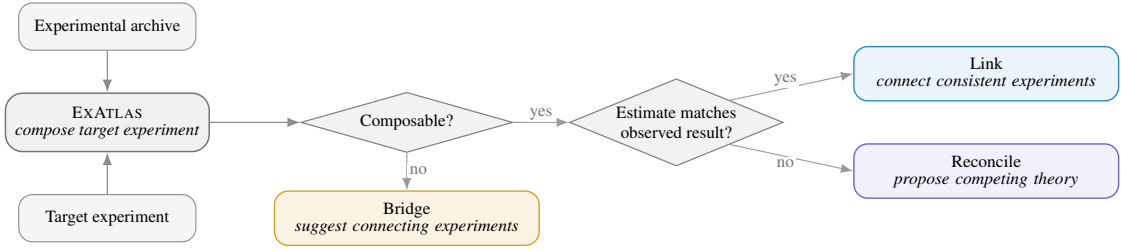

Given a target experiment and an archive of prior experiments, \ourmethod proceeds in two stages (Figure~\ref{fig:pipeline}). First, it asks whether the target's treatment and outcome can be composed from prior experiments. If so, it estimates the target's effect by composing observed effects. Second, it uses the composition to organize the target: a match links the target into the atlas, a conflict triggers a meta/competing-theory proposal, and a failure to compose triggers bridge experiments. 




\subsection{Representing Experiments}
Most experiments in social and behavioral science test causal relations between treatments and outcomes; we encode each study accordingly.
Specifically, each experiment \(i\) is encoded as \[e_i=(T_i,O_i,Z_i,\tau_i),\] where \(T_i\) is the treatment description, \(O_i\) is the outcome or dependent-variable description, \(Z_i\) contains contextual information such as experiment settings, population, and control conditions, and \(\tau_i\) is the observed treatment effect.

Because treatments and outcomes are often described briefly in papers, \ourmethod first uses an LLM (OpenAI o3 model) to enrich them into context-aware descriptions, for instance, enriching "creativity" to "employee creativity in the organization" by reading the content of the original paper. Human annotators verified 50 randomly sampled enrichments and found all of them accurate. The enriched descriptions incorporate information from \(Z_i\), which helps make composition more accurate.

We then represent the experiments in a shared treatment--outcome space. Let \(\widetilde{T}_i\) and \(\widetilde{O}_i\) denote the enriched treatment and outcome descriptions. We define
\[
x_i=
\big[
\mathrm{emb}(\widetilde{T}_i)
\,\|\,
\mathrm{emb}(\widetilde{O}_i)
\,\|\,
\mathrm{emb}(\widetilde{T}_i)\odot \mathrm{emb}(\widetilde{O}_i)
\big],
\]
where \(\mathrm{emb}(\cdot)\) denotes the text embedding function, \(\|\) denotes concatenation, and \(\odot\) denotes element-wise multiplication. The first two components encode the treatment and outcome semantics, while the interaction term captures their joint relation.



\subsection{Assessing Composability}
Given a target experiment \(e_t\), \ourmethod first asks whether \(e_t\) can be composed from prior experiments in the archive.
Because these experiments are fundamentally about causal relations between treatments and outcomes, we evaluate composability in the shared treatment--outcome space defined above. The target experiment is \textit{composable} when \(x_t\) can be closely approximated by a weighted combination of prior experiment representations.

Given a set of candidate source experiments \(\mathcal{C}(t)\) and their
reconstruction weights \(\alpha_{tj}\), we compute the distance between the
target representation and its composed approximation:
\[r_t(\alpha)
=
\left\|
x_t-\sum_{j\in\mathcal{C}(t)}\alpha_{tj}x_j
\right\|_2,\]
where \(\|\cdot\|_2\) denotes the Euclidean norm.
Because raw distances vary with archive density, we normalize the residual by the local scale:
\[
s_t=\mathrm{median}_{j\in\mathcal{C}(t)}\|x_t-x_j\|_2,
\quad
\rho_t(\alpha)=\frac{r_t(\alpha)}{s_t}.
\]

We treat \(e_t\) as composable when
\(\rho_t(\alpha) \leq \lambda\), where \(\lambda\) is a predefined threshold. This makes composition rejectable. If the target is not locally
supported, \ourmethod does not force a composition; instead, it treats the target as a gap in the atlas and uses the missing local support to guide follow-up experiment suggestions. 

We obtain \(\mathcal{C}(t)\) and \(\alpha_{tj}\) through a local search because exhaustive search over all subsets is infeasible. \ourmethod instead forms a local candidate set of up to \(K\) nearby experiments and solves the constrained optimization problem:
\[
\begin{aligned}
\alpha_t^\star
&=
\arg\min_{\alpha} r_t(\alpha),\\
&\text{subject to }
\alpha_{tj}\ge 0,\quad
\sum_{j\in\mathcal{C}(t)}\alpha_{tj}=1.
\end{aligned}
\]

\subsection{Estimating the Treatment Effect and Linking Consistent Experiments}
If a target experiment \(e_t\) is composable, \ourmethod estimates its treatment effect by applying the same weights to the observed effects of the source experiments: \(\hat{\tau}_t=
\sum_{j\in\mathcal{C}(t)}
\alpha_{tj}^\star \tau_j.
\)
Here \(\tau_j\) denotes the observed human treatment effect of source experiment
\(j\), while \(\hat{\tau}_t\) is the reconstructed estimate for the target. When
the target's human result \(\tau_t\) is available, it is used only for evaluation
and atlas organization, not for estimating \(\hat{\tau}_t\).

\ourmethod compares the direction of the composed estimate with the direction of the observed target effect. If
\(
\mathrm{sign}(\hat{\tau}_t)=\mathrm{sign}(\tau_t),
\)
then the target experiment and its source experiments are treated as a coherent local region of the atlas. In this case, \ourmethod links the target to the source experiments, indicating that the target finding can be situated within the existing experimental landscape.

\subsection{Reconciling Conflicts and Bridging Gaps}
Not every target experiment that enters the atlas becomes a link. Besides linking consistent experiments, \ourmethod provides two diagnostic outputs to support theory building and experiment design.

The first occurs when a target experiment is composable, but the composed estimate conflicts with the observed target result. In this case, experiments that are close enough to reconstruct the target nevertheless imply different effects, signaling a local theoretical tension in the atlas. \ourmethod then uses an LLM to generate a reconciliation: an explanation for why the source experiments and the target experiment may diverge despite being compositionally related. The LLM is prompted with the target experiment, the source experiments, their observed effects, and the direction of the mismatch. 
This grounds the theory proposal in the local structure of the atlas rather than in the target experiment alone.
The LLM then proposes an evidence-integration explanation, such as a meta-theory or moderator structure, that separates the conditions under which each effect obtains. 

The second diagnostic output occurs when the target is not composable. In this case, there is a gap between the target experiment and the composable region of existing literature. \ourmethod uses an LLM to suggest connecting experiments that could bridge this gap, helping turn isolated regions of the literature into a more coherent experimental landscape.
This process is iterative: an LLM proposes an experiment at each step, \ourmethod adds the proposed experiment as a hypothetical bridge to the experimental archive, and reruns the composability assessment, repeating the process until the target can be composed from existing and proposed experiments, or a step limit is reached.

We use OpenAI o3 to generate reconciliations and DeepSeek-R1 to propose connecting experiments, as both models performed better in their respective tasks; implementation details and prompts are provided in Appendices~\ref{app:implementation-details} and~\ref{app:prompts}, respectively.
\section{Validating the Composition Method}
\label{sec:theory}
Two checks support the method: a theoretical error bound shows that composition is effective when the target experiment is close to its composed approximation, and an empirical evaluation tests whether it recovers held-out experiments from the archive.

\subsection{Theoretical Justification}
When can a weighted composition of source effects recover the target effect? We show that this is possible when two conditions hold: the target is locally composable from prior experiments, and the treatment-effect surface is smooth in the neighborhood where composition occurs.


Let \(m_i\) denote the latent treatment--outcome mechanism representation of experiment \(e_i\). We assume that the treatment effect can be written as
\[
\tau_i = \mu(m_i)+\epsilon_i^{\mu},
\qquad |\epsilon_i^{\mu}|\le \delta_{\mathrm{unobs}},
\]
where \(\mu\) is a smooth treatment-effect function, \(\epsilon_i^{\mu}\) captures residual variation due to factors not represented in \(m_i\), and \(\delta_{\mathrm{unobs}}\) bounds the size of this unobserved residual.

The target experiment is locally composable if its latent representation can be approximated by a convex combination of source representations:
\[
m_t
=
\sum_{j\in\mathcal{C}(t)}
\alpha_{tj}m_j
+
\epsilon_t,
\enspace
\alpha_{tj}\ge0,
\enspace
\sum_{j\in\mathcal{C}(t)}\alpha_{tj}=1.
\]
Here \(\epsilon_t\) is the latent reconstruction residual. Small \(\|\epsilon_t\|\) means that the target lies close to the local structure induced by the source experiments.

Let
\(
\tau_t^{\mathrm{comp}}
=
\sum_{j\in\mathcal{C}(t)}
\alpha_{tj}\tau_j .
\)
Under local smoothness of the treatment-effect function \(\mu\), the error of composing source effects is bounded by
\[
\begin{aligned}
\left|
\tau_t-\tau_t^{\mathrm{comp}}
\right|
&\le
L\|\epsilon_t\|
+
\frac{H}{2}
\sum_{j\in\mathcal{C}(t)}
\alpha_{tj}
\|m_j-\bar m_t\|^2
\notag\\
&\quad+
\left|
\epsilon_t^{\mu}
-
\sum_{j\in\mathcal{C}(t)}
\alpha_{tj}\epsilon_j^{\mu}
\right|,
\end{aligned}
\]
where \(\bar m_t=\sum_{j\in\mathcal{C}(t)}\alpha_{tj}m_j\).
The bound decomposes approximation error into three interpretable components. The first term is extrapolation error: the target must be close to the composed source representation. The second term is local curvature error: composition is more accurate when the treatment-effect function is approximately linear in the local neighborhood. The third term captures the contribution of unobserved factors not represented in \(m_i\), and is bounded above by \(2\delta_{\mathrm{unobs}}\).
Full assumptions and the proof are in Appendix~\ref{app:proofs}.

In practice, the latent representation \(m_i\) is not directly available to
\ourmethod. It uses the treatment--outcome feature representation \(x_i\) as an empirical
proxy for the latent mechanism representation. The normalized residual
\(\rho_t\) estimates whether the target is close enough to the local source
composition to be treated as composable. 

\subsection{Empirical Evaluation}
\label{quant_eval}
\begin{table*}[ht]
\centering
\small
\setlength{\tabcolsep}{10pt}
\begin{tabular}{p{0.34\linewidth}cccc}
\toprule
\textbf{Method} & \textbf{Sign match \(\uparrow\)} & \textbf{MSE \(\downarrow\)} & \textbf{MAE \(\downarrow\)} & \textbf{Spearman \(\rho\) \(\uparrow\)} \\
\midrule
\rowcolor[gray]{0.95} \multicolumn{5}{c}{\textbf{Management and Psychology Papers}} \\
Direct Prediction & 71.2\% & 7.646 & 1.294 & 0.354 \\
Nearest-Experiment RAG & 78.0\% & 11.384 & 1.337 & 0.446 \\
Contributing-Experiment RAG & 84.8\% & 7.714 & 0.946 & 0.605 \\
k-NN Avg & 86.4\% & 5.220 & 1.090 & 0.630 \\
Local Ridge & 93.2\% & 4.339 & 0.994 & 0.688 \\
Synthetic Participants & 81.4\% & 7.841 & 1.690 & 0.611 \\
\ourmethod Composition & \textbf{100.0\%} & \textbf{2.772} & \textbf{0.815} & \textbf{0.830} \\

\midrule
\rowcolor[gray]{0.95} \multicolumn{5}{c}{\textbf{Education Studies}} \\
Direct Prediction & 91.0\% & 1.115 & 0.747 & 0.458 \\
Nearest-Experiment RAG & 94.9\% & 0.560 & 0.448 & 0.659 \\
Contributing-Experiment RAG & 95.7\% & 0.487 & 0.413 & 0.652 \\
k-NN Avg & \textbf{96.1\%} & 0.564 & 0.491 & 0.553 \\
Local Ridge & 95.9\% & 0.450 & 0.429 & 0.663 \\
\ourmethod Composition & 95.9\% & \textbf{0.431} & \textbf{0.394} & \textbf{0.709} \\

\bottomrule
\end{tabular}
\caption{Performance comparison on predicting treatment effects for composable experiments.}
\label{tab:results}
\end{table*}

\para{Experiment domains.} We evaluate \ourmethod across two experimental domains. The first domain, \textit{Management and Psychology Papers}, consists of 360 experiments from five management and psychology journals collected by \citet{cui2025large}. The second domain, \textit{Education Studies}, is derived from the What Works Clearinghouse study-findings archive and contains 1,580 statistically significant education-intervention findings with usable effect estimates.\footnote{https://ies.ed.gov/ncee/wwc/StudyFindings}

\para{Experimental setup.} We hold out one experiment at a time as the target and use the remaining experiments as the source archive. The composability threshold $\lambda$ determines whether a target has sufficient local support for composition. To select this threshold, we randomly hold out 20\% of the Management and Psychology Papers experiments as a validation set and select $\lambda$ by maximizing
\(
(1-\mathrm{scaled\ MSE})\times\mathrm{coverage}.
\)
Here, \(\mathrm{Coverage}\) denotes the fraction of experiments that pass the composability assessment, i.e., the fraction with \(\rho_t \le \lambda\).
The resulting threshold is \(\lambda=0.482\). 

We then fix the selected threshold and evaluate \ourmethod on the remaining 80\% of experiments. For the Education Studies domain, we use the same composability threshold and other hyperparameters without domain-specific tuning. For each domain, we report results on the subset of targets that pass the composability criterion: 59 targets for Management and Psychology Papers, and 531 targets for Education Studies. 

For each composable target, we compare the composed effect with the reported held-out effect using sign match, mean squared error (MSE), mean absolute error (MAE), and Spearman's rank correlation.
More implementation details and metric definitions are in Appendix~\ref{app:implementation-details} and \ref{app:estimator-metrics}.

\para{Baselines.}
We compare \ourmethod against six baselines, with implementation details in Appendix~\ref{app:direction-baselines}. 
(1) \textit{Direct Prediction} asks the model to predict the target effect using only the target experiment description. 
(2) \textit{Nearest-Experiment RAG} provides the model with the single most similar prior experiment retrieved by TF--IDF similarity, including its reported effect. 
(3) \textit{Contributing-Experiment RAG} provides the model with the source experiments identified by \ourmethod as contributing to the target composition, including their reported effects. 
(4) \textit{k-NN Avg} predicts the target effect by averaging the observed effects of its \(k\) nearest experiments in the same representation space used by \ourmethod, with \(k=9\), the median number of contributing experiments used by \ourmethod. 
(5) \textit{Local Ridge} fits a ridge regression model to the effects of the local neighborhood using the same representation and neighborhood size. 
(6) \textit{Synthetic Participants} uses the released LLM-as-participant replication results from \citet{cui2025large}, which simulate participant responses to the original scenarios and survey questions and calculate treatment effects by aggregating responses. Because only the \textit{Management and Psychology Papers} dataset consists of survey-based experiments with detailed questionnaire content available, this baseline is evaluated only on this domain. GPT-5.4 is used for the three prompting baselines, and Claude 3.5 Sonnet is used for the simulation baseline because it achieves the best performance in that study.


\para{Results.}
Table~\ref{tab:results} reports performance across the two experiment domains. \ourmethod achieves the strongest overall performance on both benchmarks. It recovers the reported effect direction for 100\% of targets in the \textit{Management and Psychology Papers} and 96\% of targets in the \textit{Education Studies}. It also achieves the lowest MSE and MAE and the highest Spearman correlation in both domains. These results indicate the effectiveness of the composition based on experiment embeddings.

The \textit{Contributing-Experiment RAG} baseline also performs strongly, particularly relative to the other prompting baselines. This suggests that the experiments selected by \ourmethod contain useful evidential information for predicting target effects. 
On the other hand, the stronger performance of \ourmethod over both \textit{k-NN Avg} and \textit{Local Ridge} indicates that the success depends not only on identifying a useful local neighborhood, but also on assigning composition weights to the contributing experiments. 

\section{Evaluating \ourmethod-Guided Research Directions}

\subsection{Worked Example: A Hidden $2 \times 2$ in the Hiring Literature}
\label{sec:worked-example}
Before turning to systematic evaluation, we present a real case (originally published by \citep{lee2015discrimination}) that illustrates how \ourmethod finds an anomaly and turns it into a candidate theoretical reconciliation.
In a creativity-focused, competitively framed department, a new experiment found that higher candidate age increased hiring-recommendation strength. 
The algorithm surfaced a cluster of prior findings that, together with the new finding, form a hidden \(2 \times 2\) crossing of departmental orientation (creativity vs. stability) and climate (cooperative vs. competitive). 
As shown in Table~\ref{tab:cells}, the age effect changes direction across cells. 
The contradiction does not lie in any single prior effect; it emerges from a moderator crossing that had been treated as separable in the literature.

\begin{table}[ht]
\centering
\small
\begin{tabular}{lcc}
\toprule
 & Cooperative & Competitive \\
\midrule
Creativity & $\downarrow$ (prior) & $\boldsymbol{\uparrow}$ \textbf{(new)} \\
Stability  & $\uparrow$ (prior)   & $\downarrow$ (prior) \\
\bottomrule
\end{tabular}
\caption{Sign of the age effect on hiring recommendation in the worked example. \textbf{Bold}: new finding.}
\label{tab:cells}
\end{table}

The case shows a three-step division of labor between the algorithm, the LLM, and the expert. 
The algorithm surfaces a structural anomaly in the hidden \(2 \times 2\): the effect of candidate age changes direction across cells in a way that is not explained by either departmental orientation or climate alone. 
One human expert, blind to the algorithm, did not locate this cross-cell conflict; another expert proposed industry as a moderator, which was directionally relevant but remained at single-moderator resolution and left part of the disagreement unexplained. 
Based on the anomaly, the LLM articulated the full $2 \times 2$ as \emph{Stereotype Fit--Context Alignment}, with cell-specific goals and a testable boundary condition: the reversal hinges on whether competition is framed as outcome accountability (favors age) or speed (favors youth). In \textit{post hoc} review, the experts judged the framing both plausible and useful for interpreting the conflict.
The remainder of this section systematically evaluates both diagnostic outputs of \ourmethod at scale: reconciliations when composition conflicts (\S\ref{sec:case}) and bridge suggestions when composition fails (\S\ref{sec:bridge}).

\subsection{Human Evaluation: Reconciling Conflicts}
\label{sec:case}
\begin{table}[ht]
\centering
\small
\begin{tabular}{lc}
\toprule
\textbf{Dataset} & \textbf{Agreement Rate} \\
\midrule
Management and Psychology Papers & 61.5\% \\
Education Studies & 80.6\% \\
\bottomrule
\end{tabular}
\caption{Agreement between the reconciliation direction proposed by \ourmethod and the broad direction identified by human experts.}
\label{tab:reconcile_agreement}
\end{table}

We first evaluate the Reconcile module on the two experiment domains. 
For the \textit{Management and Psychology Papers} domain, the original evaluation produced no composable-with-mismatch cases, so we relaxed the composability criterion to $\lambda' = 1.5\lambda$, and obtain 13 conflict cases for evaluating the Reconcile module. For the \textit{Education Studies} domain, we did not relax the composability criterion and obtained 12 conflict cases directly. In each case, \ourmethod instructs an LLM to propose a meta/competing theory that could reconcile the conflicting evidence. 

To evaluate these reconciliations, we recruited four Ph.D. students in social science.
Because reconciliations are difficult to evaluate with quantitative criteria, we use a qualitative expert-evaluation procedure. 
Each expert first examined the conflict case independently, without seeing the model output, and proposed their interpretation of how the findings might be reconciled. 
They were then shown the LLM-generated reconciliation and asked to compare it with their assessment.
Averaged across the two datasets, average pairwise agreement was 88.1\%, with an average Fleiss' \(\kappa\) of 0.69.

As summarized in Table~\ref{tab:reconcile_agreement}, both experts and the LLM identified the same broad direction of reconciliation in an average of 8 of 13 cases (61.5\%) for the \textit{Management and Psychology Papers} domain and 9.7 of 12 cases (80.6\%) for the \textit{Education Studies} domain.
Annotators find that the LLM often supplied a more explicit mechanism or moderator framing than the experts initially articulated. 
The worked example in \S\ref{sec:worked-example}, the age $\times$ creativity $\times$ competition case, illustrates this pattern. The \ourmethod algorithm identifies the anomaly, the LLM reconciles the conflict by articulating the moderator structure, and the experts ratify the framing in \textit{post hoc} review.
A second case involving racial-identity cues in the selection of a debate opponent follows the same pattern. One human expert wrote that ``the key advantage of LLM feedback is that it gives a more detailed view of existing theory \dots\ human rater is more limited in this sense.'' 

Human experts do not always agree with the LLM. 
In one case, an experiment found that gender in a \emph{non-empowering} coworker-assistance episode decreased self-perceived competence, echoing prior findings on \emph{empowering} assistance but reversing the direction of the effect. 
The LLM proposed a threat-benefit meta-theory to reconcile this pattern.  
A human expert disagreed, writing that this reflected the ``same mechanism, opposite types of interaction,'' and therefore added ``nothing new''  
Four additional cases follow this pattern. 
Whether a sign flip under a moderator-by-construction counts as a theoretical contribution is a methodological choice. 
These disagreements help clarify what the Reconcile module contributes. 
Reconcile does not determine whether a conflict should be treated as a new theory, a boundary condition, or a familiar mechanism instantiated in a new context.
Rather, it renders that interpretive choice explicit by transforming conflicting evidence into a theoretical proposal that experts can accept, revise, or reject. 
In this sense, Reconcile shifts the burden from implicit judgment to explicit adjudication, without claiming to settle the adjudication itself. 
We understand the framework as making precisely such choices legible. 
Taken together, the cases illustrate Reconcile’s role across modes: surfacing a cross-cell conflict in a hidden moderator structure that neither party would likely have located unaided, articulating the relevant prior literature more exhaustively than any single expert, and exposing a productive disagreement over what counts as a theoretical contribution.

\subsection{Human Evaluation: Connecting Experiments}
\label{sec:bridge}

\begin{table*}[th]
\centering
\small
\begin{tabular}{lcccccc}
\toprule
\textbf{Dimension} & \textbf{Forced Composition} & \textbf{\ourmethod Bridge} & \(\boldsymbol{t}\) & \(\boldsymbol{df}\) & \(\boldsymbol{p}\) & \textbf{95\% CI} \\
\midrule
\rowcolor[gray]{0.95} \multicolumn{7}{c}{\textbf{Management and Psychology Papers}} \\
Feasibility   & 4.89 (0.31) & 4.88 (0.33) & -0.465 & 396.78 & .642 & [-0.078, 0.048] \\
Plausibility  & 3.78 (0.94) & 3.96 (0.75) &  2.061 & 380.39 & .040 & [0.008, 0.342] \\
Connectedness & 4.10 (1.03) & 4.29 (0.75) &  2.163 & 363.05 & .031 & [0.018, 0.372] \\
\midrule
\rowcolor[gray]{0.95} \multicolumn{7}{c}{\textbf{Education Studies}} \\
Feasibility   & 4.50 (0.58) & 4.61 (0.53) & 1.989 & 395.23 & .047 & [0.001, 0.219] \\
Plausibility  & 3.79 (0.98) & 4.04 (0.49) & 3.241 & 293.32 & .001 & [0.098, 0.402] \\
Connectedness & 3.61 (0.94) & 4.14 (0.60) & 6.772 & 339.81 & $<$.001 & [0.380, 0.690] \\
\bottomrule
\end{tabular}
\caption{Comparison of Forced Composition and \ourmethod Bridge scores across three evaluation dimensions. Values for each condition report means with standard deviations in parentheses. The table also reports Welch's \(t\)-test statistics, degrees of freedom, \(p\)-values, and 95\% confidence intervals for the mean difference.}
\label{tab:baseline_new_comparison}
\end{table*}

We then conduct a human evaluation to assess whether the connecting experiments proposed by \ourmethod are useful research directions for targets that are not locally composable from the existing experimental archive. These are the targets the composability assessment rejects: their normalized reconstruction residual $\rho_t$ exceeds the threshold $\lambda$, so \ourmethod declines to compose and routes them to the Bridge module instead. 

We compare two conditions on targets that \ourmethod identifies as non-composable. The first, Forced Composition, runs the composition algorithm on each target regardless of whether the composability assessment rejects it, returning for each target a list of contributing experiments from the existing archive. This represents a direct-composition strategy without a refusal or gap-detection mechanism. The second, \ourmethod Bridge, uses the proposed gap-bridging procedure: rather than forcing a composition under insufficient support, \ourmethod generates one or more connecting experiments intended to link the target to nearby experimental evidence, until the target becomes composable from existing and proposed experiments or a step limit is reached. 
This could also be viewed as the ablation study of the Bridge module.

\para{Evaluation dimensions.}
Each experiment (contributing experiment in Forced Composition and connecting experiment in \ourmethod Bridge) is evaluated independently along three dimensions using a five-point Likert scale: \textbf{Feasibility}, \textbf{Plausibility}, and \textbf{Connectedness}. 
Feasibility captures whether the experiment can be empirically implemented with a clear design, including a manipulable independent variable, a measurable dependent variable, and a plausible data collection procedure.
Plausibility assesses whether the proposed relationship between the independent and dependent variables is theoretically, empirically, or logically reasonable.
Connectedness evaluates whether the suggested experiment meaningfully links or explains the original input experiments, beyond surface-level keyword overlap.
Six raters (three Ph.D. students and three advanced undergraduates, all with backgrounds in organizational behavior or industrial-organizational psychology) evaluated 100 lists of contributing experiments per condition (Forced Composition vs. \ourmethod Bridge) in each domain on the three dimensions. 
Each experiment received at least two independent ratings. 
Full assignment and demographic details are in Appendix~\ref{app:human_eval}.


\para{Inter-Rater agreement.}
Inter-rater reliability was moderate to high across conditions (ICC(1,2) range 0.45--0.90), and within-one agreement exceeded 0.97 on every dimension (Table~\ref{tab:judge_agreement} in Appendix~\ref{app:human_eval}). 
Feasibility ratings were near ceiling in both conditions, likely attenuating ICC estimates for this dimension. 
We therefore focus substantive comparisons on Plausibility and Connectedness.

\para{Comparison results.}
We conducted Welch's independent-samples \(t\)-tests as appropriate to compare the two conditions across the three rating dimensions.
Welch's \(t\)-test was used because it does not assume equal variances across groups and is therefore more appropriate when the rating distributions differ in variance across methods.
As shown in Table~\ref{tab:baseline_new_comparison}, \ourmethod Bridge received higher average ratings than Forced Composition on all three dimensions. \ourmethod Bridge scored 4.75 versus 4.70 for Feasibility, 4.00 versus 3.79 for Plausibility, and 4.22 versus 3.86 for Connectedness. The advantage was most consistent for Plausibility and Connectedness: both dimensions were significantly higher for \ourmethod Bridge. The evidence for Feasibility was weaker and less significant, suggesting that \ourmethod Bridge does not gain its advantage by proposing experiments that are harder or easier or clearer to implement.

Together, these results support the gap-bridging design of \ourmethod. 
\ourmethod Bridge proposes new experiments that help reduce the gap between the target and nearby experimental evidence. 
Importantly, the proposed experiments remain empirically grounded: compared with the published experiments retrieved by Forced Composition, they are equally feasible and more plausible.

\section{Conclusion}

This paper introduces \ourmethod, a framework for turning fragmented experimental archives into an atlas of links, conflicts, and gaps. 
Given a target experiment, it asks whether the archive of human experiments is close enough, in treatment and outcome, to compose a prediction. When it is, \ourmethod links consistent evidence or turns mismatches into theoretical reconciliation. 
When it is not, \ourmethod declines to compose and proposes bridge experiments intended to make composition possible. 
On the targets it certifies as locally supported, predicted directions match observed ones in 98\% cases. 
On targets it declines, human raters judge the proposed bridge experiments to be feasible, plausible, and well connected to the target literature. 

Read compositionally, an experimental archive becomes a resource for assessing where its own evidence reaches, where it falls short, and where those limits point to meaningful research directions.
If social science is to escape what \citet{watts2017should} called the incoherency problem, it needs infrastructure for reading the archive as a map rather than a pile. \ourmethod is one such infrastructure.


\section*{Limitations}



\textbf{The archive is itself biased.} What gets published, what gets coded as a clean treatment-outcome experiment, and what gets a coded effect direction all reflect editorial and disciplinary selection. Local composition inherits these biases; the atlas is dense where the literature is dense and silent where it is silent.


\textbf{The meta-competing-theory reconciliation module is a generator, not a verifier}. Section 5.2 shows that the LLM can articulate moderator structures that an expert ratifies, but ratification is not validation. The proper next step is to run the boundary-condition experiments that the module proposes.

\section*{Ethics Statement}
The research ideas, bridge experiments, and theoretical reconciliations generated by LLMs may reflect biases present in their training data and may unintentionally resemble existing work without proper attribution. 
They should therefore be treated as candidate research directions rather than validated scientific claims, and should not be adopted for practical use without careful expert review, literature checking, and empirical validation. 
Because the framework operates on human-subject experimental evidence, researchers should also consider whether proposed experiments involve sensitive populations, identity-related manipulations, or other ethical risks before implementation. 
Any use of LLM-generated ideas or explanations should be disclosed transparently to preserve research integrity. In this paper, we only used AI to polish text and write code. All AI-generated content has been well examined by authors before adoption.


\bibliography{references}

\appendix
\section{Details of the Theoretical Justification}
\label{app:proofs}

\subsection{Assumptions}

We state the assumptions used in the theoretical analysis of \ourmethod. 

\paragraph{A1. Source internal validity.}
For each source experiment \(e_i\), random assignment makes the treatment and control groups comparable, so the difference in their average outcomes estimates the treatment effect \(\tau_i\).
That is, treatment assignment \(A_i\in\{0,1\}\) is randomized:
\[
A_i \perp (Y_i(1),Y_i(0)).
\]
The observed outcome satisfies consistency:
\[
Y_i = A_iY_i(1)+(1-A_i)Y_i(0).
\]
Therefore, the source average treatment effect is identified by
\[
\tau_i
=
\mathbb{E}[Y_i\mid A_i=1]
-
\mathbb{E}[Y_i\mid A_i=0].
\]

\paragraph{A2. Approximate mechanism sufficiency.}
Each experiment \(e_i\) has a latent treatment--outcome representation
\[
m_i=m(T_i,O_i,Z_i)\in\mathbb{R}^d,
\]
which captures the treatment, outcome, and contextual features that are relevant for treatment-effect variation across experiments. 
We assume the treatment effect can be written as
\[
\tau_i=\mu(m_i)+\epsilon_i^{\mu},
\qquad |\epsilon_i^{\mu}|\leq \delta_{\mathrm{unobs}},
\]
where \(\mu\) maps mechanism representations to treatment effects, and \(\epsilon_i^{\mu}\) captures residual variation from features not represented in \(m_i\). 
The bound \(\delta_{\mathrm{unobs}}\) limits the contribution of such unobserved factors.


\paragraph{A3. Local smoothness.}
The treatment-effect function \(\mu\) is locally smooth around the target and its selected source experiments. 
Specifically, we assume a local Lipschitz condition,
\[
|\mu(m)-\mu(m')|\le L\|m-m'\|,
\]
and bounded local curvature,
\[
\|\nabla^2\mu(m)\|_{\mathrm{op}}\le H,
\]
for experiments in this neighborhood. 
Intuitively, nearby experiments in mechanism space should have related treatment effects, and the effect surface should not change abruptly within the local neighborhood.

\paragraph{A4. Local support.}
The target representation lies approximately in the convex hull of nearby source
representations:
\[
\begin{aligned}
m_t&=\sum_{j\in\mathcal{C}(t)}\alpha_{tj}m_j+\epsilon_t,\\
\alpha_{tj}&\ge 0,\qquad
\sum_{j\in\mathcal{C}(t)}\alpha_{tj}=1.
\end{aligned}
\]
The residual \(\epsilon_t\) measures the degree to which the target lies outside
local source support.

\paragraph{A5. Representation validity.}
The latent mechanism representation \(m_i\) is not observed directly. 
In practice, \ourmethod uses the text-based experiment representation \(x_i\), constructed from the treatment and outcome descriptions, as an empirical proxy. 
We assume that this representation approximately preserves local neighborhood structure, so that nearby experiments and reconstruction weights in \(x\)-space are informative about the corresponding relationships in latent mechanism space.

\subsection{Proof of the Error Bound}
Let
\[
\bar m_t
=
\sum_{j\in\mathcal{C}(t)}
\alpha_{tj}m_j .
\]
By local compositional support,
\[
m_t=\bar m_t+\epsilon_t.
\]
The target effect is
\[
\tau_t=\mu(m_t)+\epsilon_t^{\mu},
\]
and the composed effect is
\[
\tau_t^{\mathrm{comp}}
=
\sum_{j\in\mathcal{C}(t)}
\alpha_{tj}\tau_j .
\]
Using \(\tau_j=\mu(m_j)+\epsilon_j^{\mu}\), we have
\[
\tau_t^{\mathrm{comp}}
=
\sum_{j\in\mathcal{C}(t)}
\alpha_{tj}\mu(m_j)
+
\sum_{j\in\mathcal{C}(t)}
\alpha_{tj}\epsilon_j^{\mu}.
\]

We decompose the approximation error:
\[
\begin{aligned}
&|\tau_t-\tau_t^{\mathrm{comp}}| =
\\
&\quad
\Bigl|
\mu(m_t)
-\!\sum_{j\in\mathcal{C}(t)}\!\alpha_{tj}\mu(m_j)
+\epsilon_t^{\mu}
-\!\sum_{j\in\mathcal{C}(t)}\!\alpha_{tj}\epsilon_j^{\mu}
\Bigr|
\\
&\le
\Bigl|
\mu(m_t)-\!\sum_{j\in\mathcal{C}(t)}\!\alpha_{tj}\mu(m_j)
\Bigr|
+
\Bigl|
\epsilon_t^{\mu}
-\!\sum_{j\in\mathcal{C}(t)}\!\alpha_{tj}\epsilon_j^{\mu}
\Bigr|.
\end{aligned}
\]

For the mechanism component, add and subtract \(\mu(\bar m_t)\):
\[
\begin{aligned}
&\left|
\mu(m_t)
-
\sum_{j\in\mathcal{C}(t)}
\alpha_{tj}\mu(m_j)
\right|
\notag\\
&\quad\le
\left|
\mu(m_t)-\mu(\bar m_t)
\right|
\notag\\
&\qquad+
\left|
\mu(\bar m_t)
-
\sum_{j\in\mathcal{C}(t)}
\alpha_{tj}\mu(m_j)
\right|.
\end{aligned}
\]

The first term is controlled by local Lipschitz smoothness:
\[
\left|
\mu(m_t)-\mu(\bar m_t)
\right|
\le
L\|m_t-\bar m_t\|
=
L\|\epsilon_t\|.
\]

We now bound the second term. 
For each source experiment \(e_j\), apply Taylor's theorem to \(\mu(m_j)\) around \(\bar m_t\):
\[
\mu(m_j)
=
\mu(\bar m_t)
+
\nabla\mu(\bar m_t)^\top(m_j-\bar m_t)
+
R_j,
\]
where the second-order remainder satisfies
\[
|R_j|
\le
\frac{H}{2}
\|m_j-\bar m_t\|^2
\]
because \(\|\nabla^2\mu(m)\|_{\mathrm{op}}\le H\) in the local neighborhood.

Taking the weighted sum gives
\[
\begin{aligned}
\sum_{j\in\mathcal{C}(t)}
\alpha_{tj}\mu(m_j)
&=
\sum_{j\in\mathcal{C}(t)}
\alpha_{tj}\mu(\bar m_t) \\
&{}+
\nabla\mu(\bar m_t)^\top
\sum_{j\in\mathcal{C}(t)}
\alpha_{tj}(m_j-\bar m_t) \\
&{}+
\sum_{j\in\mathcal{C}(t)}
\alpha_{tj}R_j .
\end{aligned}
\]
Since the weights sum to one,
\[
\sum_{j\in\mathcal{C}(t)}
\alpha_{tj}\mu(\bar m_t)
=
\mu(\bar m_t).
\]
The first-order term cancels because
\[
\begin{aligned}
&
\sum_{j\in\mathcal{C}(t)}
\alpha_{tj}(m_j-\bar m_t)
\notag\\
&\quad=
\sum_{j\in\mathcal{C}(t)}
\alpha_{tj}m_j
-
\bar m_t
\sum_{j\in\mathcal{C}(t)}
\alpha_{tj}
\notag\\
&\quad=
\bar m_t-\bar m_t
=
0.
\end{aligned}
\]
Thus,
\[
\sum_{j\in\mathcal{C}(t)}
\alpha_{tj}\mu(m_j)
=
\mu(\bar m_t)
+
\sum_{j\in\mathcal{C}(t)}
\alpha_{tj}R_j .
\]
Therefore,
\[
\begin{aligned}
\left|
\mu(\bar m_t)
-
\sum_{j\in\mathcal{C}(t)}
\alpha_{tj}\mu(m_j)
\right|
&=
\left|
\sum_{j\in\mathcal{C}(t)}
\alpha_{tj}R_j
\right| \\
&\hspace{-3em}\le
\sum_{j\in\mathcal{C}(t)}
\alpha_{tj}|R_j| \\
&\hspace{-3em}\le
\frac{H}{2}
\sum_{j\in\mathcal{C}(t)}
\alpha_{tj}
\|m_j-\bar m_t\|^2 .
\end{aligned}
\]

Combining the two bounds gives
\[
\begin{aligned}
\left|
\tau_t-\tau_t^{\mathrm{comp}}
\right|
&\le
L\|\epsilon_t\|
+
\frac{H}{2}
\sum_{j\in\mathcal{C}(t)}
\alpha_{tj}
\|m_j-\bar m_t\|^2 \\
&\quad+
\left|
\epsilon_t^{\mu}
-
\sum_{j\in\mathcal{C}(t)}
\alpha_{tj}\epsilon_j^{\mu}
\right|.
\end{aligned}
\]
Since \(|\epsilon_i^{\mu}|\le \delta_{\mathrm{unobs}}\) and the weights sum to one, the residual term is bounded by \(2\delta_{\mathrm{unobs}}\).

\section{Implementation Details of \ourmethod}
\label{app:implementation-details}




In the composability assessment, the embeddings are computed with
\texttt{sentence-transformers/all-mpnet-base-v2}, an MPNet-based sentence
embedding model. For each target, \ourmethod first computes distances to all candidate source experiments in the representation space. It then forms a local candidate set by taking experiments whose distance is within \(1.5\) times the median candidate distance, capped at \(K=30\) nearest candidates for scalability. 

Within this selected neighborhood, \ourmethod solves the nonnegative, unit-sum reconstruction problem, adding a ridge penalty
\(10^{-2}\|\alpha_t\|_2^2\) for numerical stability. The implementation uses
OSQP when available and falls back to the default convex solver otherwise. If
optimization fails to return weights, uniform weights over the selected
neighbors are used as a conservative fallback.



We use the o3 model for the reconciliation step. When the composed direction derived from contributing experiments conflicts with the observed sign in human experiments, we prompt the model to propose a reconciliation (e.g., a meta theory) to account for the discrepancy (see Appendix \ref{app:prompts} for the prompt). 

\begin{figure}[h]
    \includegraphics[width=0.3\textwidth]{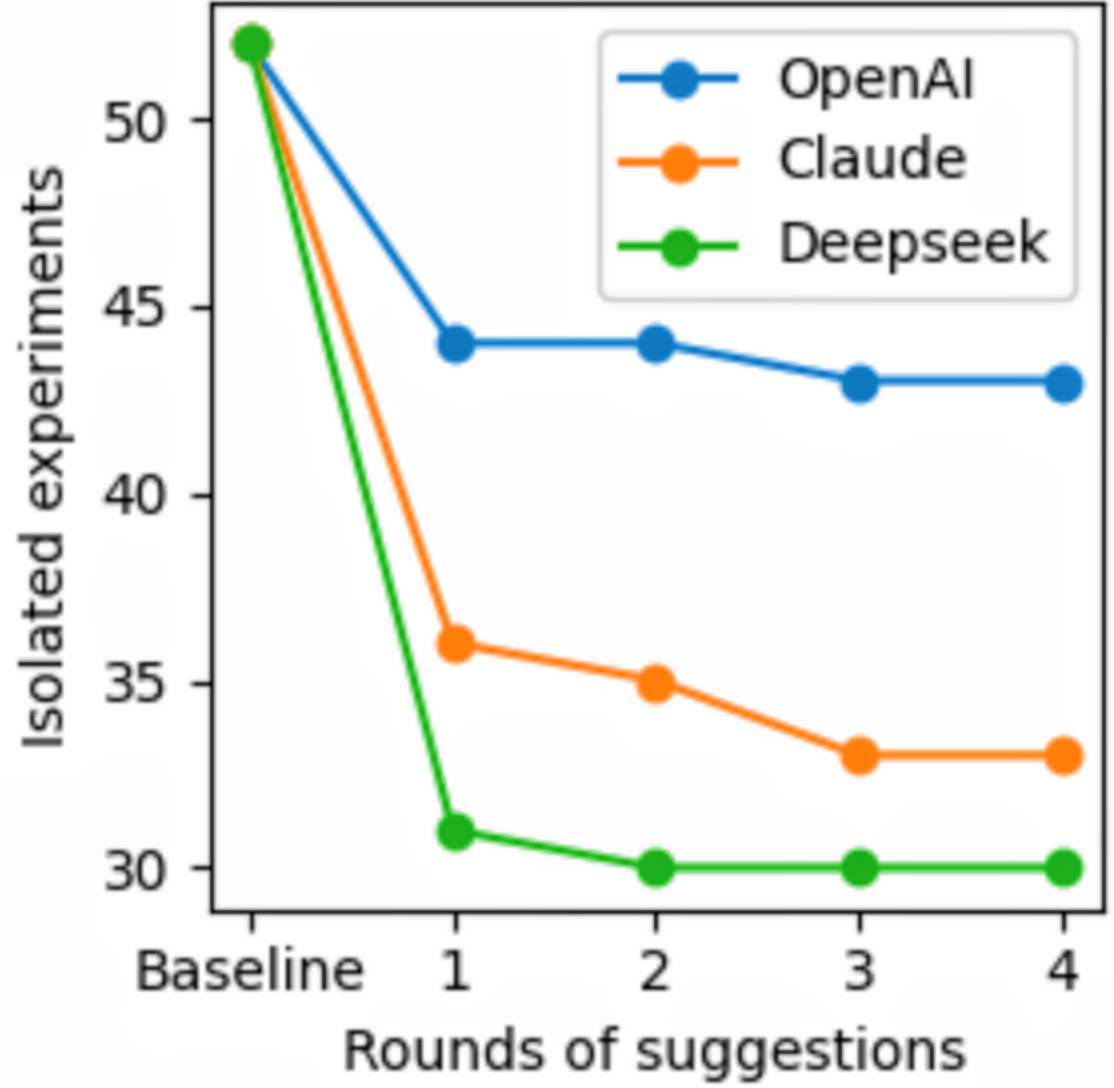}
    \centering
    \caption{DeepSeek-R1 dominates the arena when it comes to suggesting new experiments to connect existing ones (vs. OpenAI o3 and Claude Opus 4.5).}
    \label{fig:rounds}
    \vspace{-4mm}
\end{figure}

For the bridging step, we follow a stepwise search process. At each step, we rerun the algorithm jointly over the existing experiments and the newly suggested ones, checking whether the ratio of isolated experiments decreases. We evaluated several advanced models for proposing connecting experiments and found that DeepSeek-R1 performed best (Figure \ref{fig:rounds}). The first round connects nearly all of the isolated experiments\footnote{"isolate" means that they can neither contribute to other target experiments nor serve as a target themselves; thus, there are fewer isolated experiments than those that are rejected in composition and moved to the reconcile and bridge modules.} that can be connected; subsequent rounds yield diminishing returns. We attribute this to the nature of our dataset—the collection is incomplete, so some experiments are genuinely unlinkable. We thus show the results in the main paper based on one round of suggestions proposed by Deepseek-R1.

\section{Evaluation Metrics for the Composition Experiment}
\label{app:estimator-metrics}

We evaluate predicted treatment effects using four metrics: sign match, MSE, MAE, and Spearman's \(\rho\). 
Let \(\hat{\tau}_t\) denote the predicted treatment effect for target experiment \(t\), and let \(\tau_t\) denote the reported human treatment effect. 

We define sign match as the fraction of targets for which the predicted effect has the same direction as the reported human effect:
\[
\mathrm{SignMatch}
=
\frac{1}{M}
\sum_{t=1}^{M}
\mathbf{1}\!\left[
\mathrm{sign}(\hat{\tau}_t)
=
\mathrm{sign}(\tau_t)
\right],
\]
where \(M\) is the number of evaluated target experiments. 

Let
\(
e_t=\hat{\tau}_t-\tau_t
\)
denote the prediction error. 
We report mean squared error:
\[
\mathrm{MSE}
=
\frac{1}{M}
\sum_{t=1}^{M}
e_t^2,
\]
and mean absolute error:
\[
\mathrm{MAE}
=
\frac{1}{M}
\sum_{t=1}^{M}
|e_t|.
\]

Finally, we report Spearman's rank correlation between the predicted effects and the reported human effects. 
This metric evaluates whether the method preserves the relative ordering of treatment effects across target experiments.

\section{Baselines for the Composition Experiment}
\label{app:direction-baselines}


The first three baselines ask an LLM to predict the target effect size and direction.

\paragraph{Direct Prediction.}
The Direct Prediction baseline uses only the target experiment information, including the treatment, outcome, and detailed conditions of the experiment. 

\paragraph{Nearest-Experiment RAG.}
The Nearest-Experiment RAG baseline augments the target description with one retrieved prior experiment. 
The retrieved experiment is selected by TF--IDF cosine similarity over the experiment descriptions. 
The prompt includes the retrieved experiment's reported effect, allowing the model to use a simple nearest-neighbor evidence signal.

\paragraph{Contributing-Experiment RAG.}
The Contributing-Experiment RAG baseline augments the target description with the contributing experiments identified by \ourmethod. 
Due to length limit, at most 20 contributing experiments are included, ordered by their weights. 

\paragraph{k-NN Avg.}
For each target experiment, we compute cosine similarity between the target and every other candidate experiment in the archive, excluding the target's own group (i.e., other findings from the same study), and average the observed effects of the $k=9$ most similar candidates.

\paragraph{Local Ridge.}
Using the same $k=9$ nearest neighbors, we fit an $\ell_2$-regularized linear regression (ridge, penalty $\alpha=1$) mapping each neighbor's representation vector to its observed effect, then apply the fitted model to the target's own representation vector to obtain its predicted effect. This differs from k-NN Avg in that neighbors are weighted implicitly through the regression fit rather than averaged uniformly, and the model can extrapolate outside the range of the neighbors' observed effects. 

\paragraph{Synthetic Participants.}
The Synthetic Participants baseline uses the released LLM-as-participant replication results from \citet{cui2025large}, where an LLM simulates human participants given the questionnaires in the treatment and control conditions. 
For each target experiment, simulated responses are aggregated within treatment and control conditions.

\section{Human Evaluation of Connecting Experiments}
\label{app:human_eval}
We conducted a human evaluation to assess the quality of the generated experiment suggestions, following established practices for evaluating automatically generated text \citep{howcroft-etal-2020-twenty,van-der-lee-etal-2019-best}. 
We evaluated suggestions along three dimensions: Feasibility, Plausibility, and Connectedness, drawing on prior work on idea quality and subjective assessment of creative products \citep{dean_etal_2006, amabile_1982}. 

Six independent raters participated in the evaluation, including three Ph.D. students and three advanced undergraduate students, all of whom had backgrounds in organizational behavior, industrial-organizational psychology, or closely related areas. 
The raters included two men and four women.

Raters were randomly assigned to one of two conditions. 
Three raters evaluated the 100 Forced Composition suggestions, and the remaining three raters evaluated the 100 \ourmethod Bridge suggestions. 
Within each condition, study assignments were organized such that every experiment was evaluated by at least two raters.

For each annotation task, raters were shown a experiment to evaluate (contributing experiment in Forced Composition and generated connecting experiment in \ourmethod Bridge) together with its source context. They then independently evaluated the experiment on a five-point Likert scale. 
Raters were instructed to assess each dimension independently, as a experiment could score highly on one dimension while scoring lower on another. 

\paragraph{Feasibility.}
Feasibility captures the extent to which the experiment can be operationalized in practice. A highly feasible experiment is one for which a concrete empirical design, including data collection, manipulation or intervention, experimental procedure, and measurement strategy, can be readily specified and executed.

\textit{Item:} How feasible is it to empirically implement the  experiment?

\begin{itemize}
    \item 1 -- Not feasible: Cannot be realistically operationalized or tested.
    \item 2 -- Low feasibility: Presents significant practical or methodological challenges.
    \item 3 -- Moderate feasibility: Feasible with some effort or additional assumptions.
    \item 4 -- High feasibility: Can be implemented using standard research methods.
    \item 5 -- Very high feasibility: Easily operationalized with a clear and straightforward empirical design.
\end{itemize}

\paragraph{Plausibility.}
Plausibility captures the likelihood that a theoretically grounded, meaningful relationship exists between the independent and dependent variables. 
This dimension reflects whether the proposed relationship is supported by existing theoretical or empirical knowledge, as well as by the internal logic of the experiment.

\textit{Item:} How plausible is the relationship between the  independent and dependent variables of the experiment?

\begin{itemize}
    \item 1 -- Not plausible: The relationship is unlikely or lacks logical or theoretical support.
    \item 2 -- Slightly plausible: The relationship has weak or unclear justification.
    \item 3 -- Moderately plausible: The relationship has some reasonable basis, but remains uncertain.
    \item 4 -- Very plausible: The relationship has strong theoretical or intuitive support.
    \item 5 -- Highly plausible: The relationship is very likely and well supported by existing knowledge or logic.
\end{itemize}

\paragraph{Connectedness.}
Connectedness captures the degree to which the experiment helps explain, integrate, or bridge the original input experiments. A highly connected experiment provides a meaningful link between the input experiments and clarifies how they may be related, regardless of the direction or sign of the relationship.

\textit{Item:} To what extent does the experiment help connect or explain the original input experiments?

\begin{itemize}
    \item 1 -- Not connected: Does not meaningfully relate to or explain the input experiments.
    \item 2 -- Weakly connected: Only superficially relates to the input experiments.
    \item 3 -- Moderately connected: Has some relevance, but limited explanatory value.
    \item 4 -- Strongly connected: Clearly helps explain, extend, or relate the input experiments.
    \item 5 -- Highly connected: Provides a strong and meaningful bridge that integrates or explains the input experiments.
\end{itemize}
\begin{table*}[ht]
\centering
\small
\setlength{\tabcolsep}{6pt}
\begin{tabular}{@{}llcccc@{}}
\toprule
Condition & Dimension & ICC(1,2) & Within-one agreement & Mean abs. diff & \% at scale max \\
\midrule
\rowcolor[gray]{0.95} \multicolumn{6}{c}{\textbf{Management and Psychology Papers}} \\
& Feasibility  & 0.45 & 1.00 & 0.14 & 0.89 \\
Forced Composition & Plausibility   & 0.83 & 0.97 & 0.42 & 0.24 \\
& Connectedness  & 0.87 & 0.97 & 0.42 & 0.45 \\
\midrule
& Feasibility  & 0.58 & 1.00 & 0.13 & 0.88 \\
\ourmethod Bridge & Plausibility   & 0.75 & 0.97 & 0.39 & 0.24 \\
& Connectedness  & 0.69 & 0.98 & 0.49 & 0.45 \\
\midrule
\rowcolor[gray]{0.95} \multicolumn{6}{c}{\textbf{Education Studies}} \\
& Feasibility  & 0.67 & 1.00 & 0.33 & 0.53 \\
Forced Composition & Plausibility  & 0.90 & 0.99 & 0.34 & 0.20 \\
& Connectedness  & 0.86 & 0.99 & 0.42 & 0.165 \\
\midrule
& Feasibility  & 0.71 & 1.00 & 0.25 & 0.625 \\
\ourmethod Bridge & Plausibility  & 0.74 & 1.00 & 0.20 & 0.13 \\
& Connectedness  & 0.47 & 0.99 & 0.49 & 0.265 \\
\bottomrule
\end{tabular}
\caption{Agreement statistics for Forced Composition and \ourmethod Bridge across three evaluation dimensions. ICC(1,2) measures inter-rater reliability, within-one agreement reports percentage within one point, mean absolute difference reports average score disagreement, and \% at scale max reports percentage at the maximum scale value.}
\label{tab:judge_agreement}
\end{table*}

\section{Prompts}
\label{app:prompts}


\lstdefinestyle{promptstyle}{
    basicstyle=\ttfamily\footnotesize,
    breaklines=true,
    breakatwhitespace=false,
    columns=fullflexible,
    keepspaces=true,
    showstringspaces=false,
    frame=single,
    framesep=6pt,
    rulecolor=\color{black!40},
    backgroundcolor=\color{black!3},
    xleftmargin=0pt,
    xrightmargin=0pt,
    aboveskip=8pt,
    belowskip=8pt,
    literate=%
        {→}{$\rightarrow$}{1}
        {“}{``}{1}
        {“}{‘’}{1}
        {‘}{`}{1}
        {‘}{‘}{1}
}

This section documents the verbatim prompts used in our pipeline. Placeholders enclosed in curly braces (e.g., \texttt{\{iv\}}, \texttt{\{dv\}}) are dynamically substituted at runtime with the corresponding field values from each experimental record.

\subsection{Bridge-Experiment Generation Prompt}
\label{sec:prompts:bridge}

This prompt asks the model to propose \emph{bridge experiments} that logically connect a target experiment to an existing body of literature, without duplicating any experiment already known to the system.

\begin{lstlisting}[style=promptstyle]
You are a professor in management and psychology research domain. You mainly use experimental methods. Your task is to bridge the gap between a *target experiment* and the existing body of scientific literature to make the whole theory space more connected and self-contained.

[Target Experiment]:
"How does {IV} impact {DV}?"

[Context]:
The target experiment currently lacks strong connections to established research or prior experimental evidence. Thus, it cannot be logically derived from the collection of existing body of literature. Relevant existing body of literature (separated by '; ') is listed below:
"{literature}"

[Task]:
Propose bridge experiments that logically connect the target experiment to the existing body of literature, such that the derived synergy of prior evidence and the newly proposed experiments form a coherent and cumulative pathway toward linking, justifying and motivating the target experiment.

[Requirements for each proposed experiment]:
1. Try to state the proposed experiment in the **simple form of independent/dependent variables and their relations** (e.g., **Variable A positively/negatively impacts variable B.**).
2. You should try to **avoid** introducing complex conditioning, mediation, or moderation relations between dependent and independent variables unless doing so is necessary to establish a meaningful link between experiments.
3. <**Avoid** introducing theories/experiments that directly compete with or contradict> the existing evidence or the derived synergy of existing evidence.
4. Ensure the proposed experiments are **logically coherent and empirically testable**.
5. Embed **concrete details** into each proposed experiment.
6. **Be concise and creative**.
7. You should select the number of proposed experiments that is **the most appropriate** for the connection to the space of the established experiments.

You should **avoid** proposing experiments that duplicate any of those in the given list below:
"{listofknown}"

Return only the proposed experiments.
If more than one experiment is proposed, **separate them with ";"**
\end{lstlisting}

\subsection{Conflict-Reconciliation Prompts}
\label{sec:prompts:reconcile}

A reconciliation prompt is used to probe whether an observed target finding contradicts prior literature, and, if so, to elicit a unified meta/competing-theoretical account.

A mismatch between the composed and observed effect directions identifies a candidate conflict. We then apply the consistency check in Q1 to assess whether the observed target finding contradicts the source findings or the hypotheses they jointly imply. Only cases receiving a ``Yes'' response proceed to reconciliation in Q2 and enter the human evaluation.

\begin{lstlisting}[style=promptstyle]
I ran an experiment on humans and found:
{Observed target result}

Prior literature suggests:
{Results of contributing experiments}

Answer the following:
- Q1 - Consistency check:
 Do my findings contradict the prior findings or the hypotheses implied by them? Answer Yes or No.

- Q2: If Yes, reconcile the conflict.
 Source of discrepancy: Identify concrete reasons for the conflict (e.g., differences in context, moderator conditions, populations, operationalization of variables, model assumptions, or distinct variants within a broader meta-theory).
 Unified explanation: Propose a single, coherent reconciliation (e.g., a meta theory) that can accommodate both my findings and the prior literature. Be explicit about the mechanisms or conditions under which each result holds.

Be creative. Skip Q2 if the answer to Q1 is No.
\end{lstlisting}

\end{document}